\documentclass[prl,twocolumn,showpacs]{revtex4}
\usepackage{amsmath,amssymb,subfigure}
\usepackage[pdftex]{graphicx}
\usepackage{color}

\providecommand{\myheading}[1]{\textbf{#1}}   
\providecommand{\myheading}[1]{           }   

\providecommand{\rrr}{\mathbf{r}}
\providecommand{\SSS}{\mathbf{S}}
\providecommand{\cccc}{c^{\phantom\dag}}
\providecommand{\cdag}{c^\dag}

\begin{document}

\title{Fermions in 3D Optical Lattices: Cooling Protocol to Obtain Antiferromagnetism}
\author{Thereza Paiva$^{1}$, Yen Lee Loh$^2$, Mohit Randeria$^{2}$, Richard T. Scalettar$^3$, and Nandini Trivedi$^2$}
\affiliation{
$^{1}$
Instituto de Fisica, Universidade Federal do Rio de
Janeiro Cx.P. 68.528, 21941-972 Rio de Janeiro RJ, Brazil\\
$^{2}$Department of Physics, The Ohio State University, Columbus, OH
43210, USA \\
$^{3}$Department of Physics, University of California, Davis, CA 95616, USA
}

\begin{abstract}
A major challenge in realizing antiferromagnetic (AF) and
superfluid phases in optical lattices is the ability to cool fermions.
We determine the equation of state for the 3D repulsive Fermi-Hubbard model
as a function of the chemical potential, temperature and repulsion
using unbiased determinantal quantum Monte Carlo methods,
and we then use the local density approximation to model a harmonic trap.
We show that increasing repulsion leads to cooling, but \emph{only in a trap},
due to the redistribution of entropy from the center to the metallic wings.
Thus, even when the average entropy per particle is {\em larger} than that required for antiferromagnetism in the homogeneous system, the trap enables the formation of an AF Mott phase.
\end{abstract}

\maketitle

\myheading{Introduction:}
One of the most exciting themes in condensed matter physics is how complex states of matter emerge from simple Hamiltonians. 
In particular, the repulsive Fermi-Hubbard model gives rise to a rich variety of behavior, including a Mott insulating regime, an antiferromagnetically ordered N\'eel state, and possibly a ``high-temperature'' $d$-wave superfluid.

Cold atomic gases are unique in being clean and tunable systems that offer tremendous promise for exploring such Hamiltonians.
The Fermi-Hubbard model can be emulated using an optical lattice with two hyperfine species of fermions~\cite{esslingerReview}. 
Several experimental feats have already been accomplished:
the observation of sharp Fermi surfaces for free fermions in an optical lattice~\cite{kohl2005},
and of the Mott insulating regime for repulsively interacting fermions~\cite{jordens2008,schneider2008}. 
The next step in this quest is to go to even lower temperatures, where the local moments order to form a N\'eel antiferromagnet.

	\begin{figure}[!htb]
	\subfigure{
		\includegraphics[width=0.99\columnwidth]{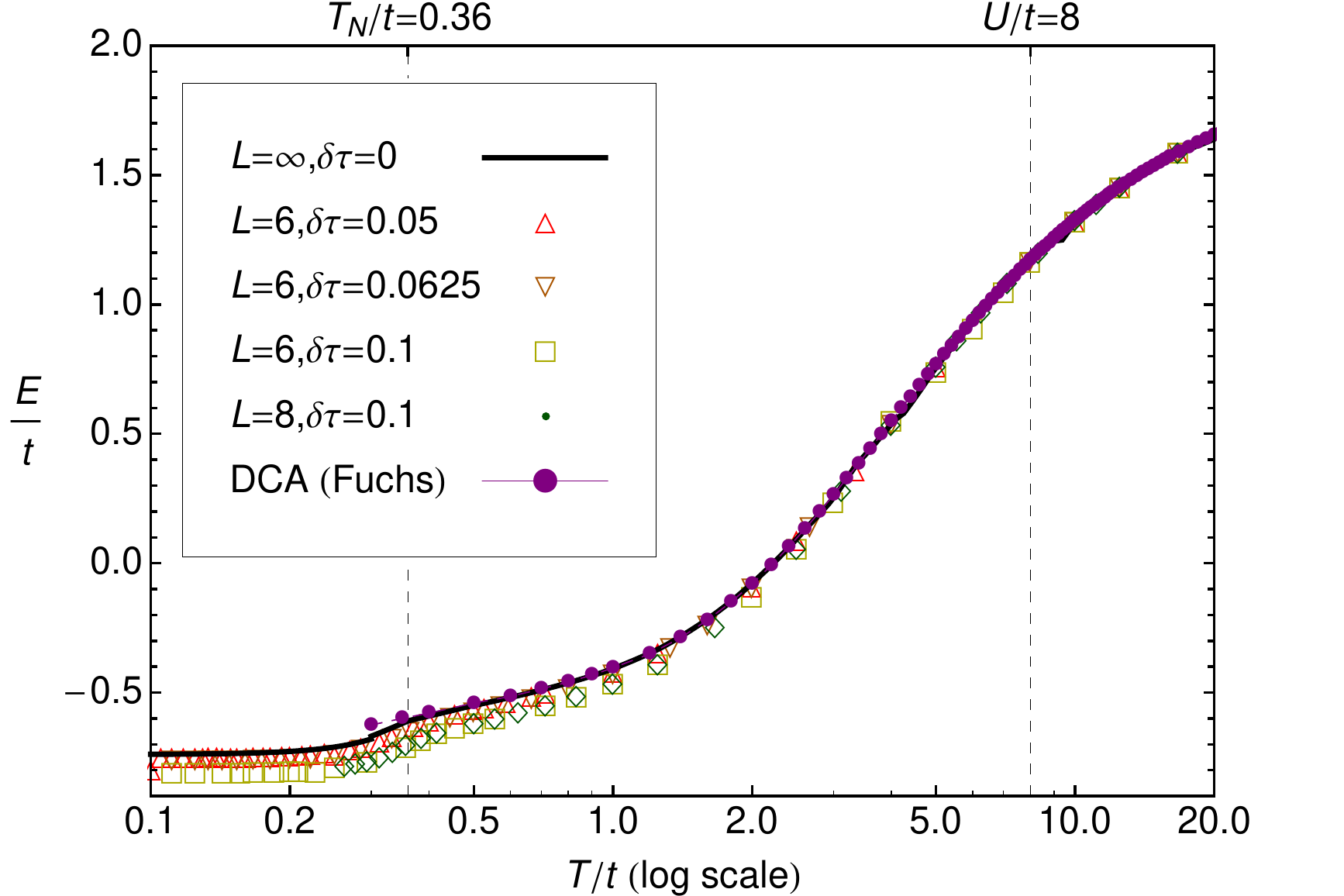}
		\label{ETFinal}
	}
	\vspace{-1cm}
	\subfigure{
		\includegraphics[width=0.99\columnwidth]{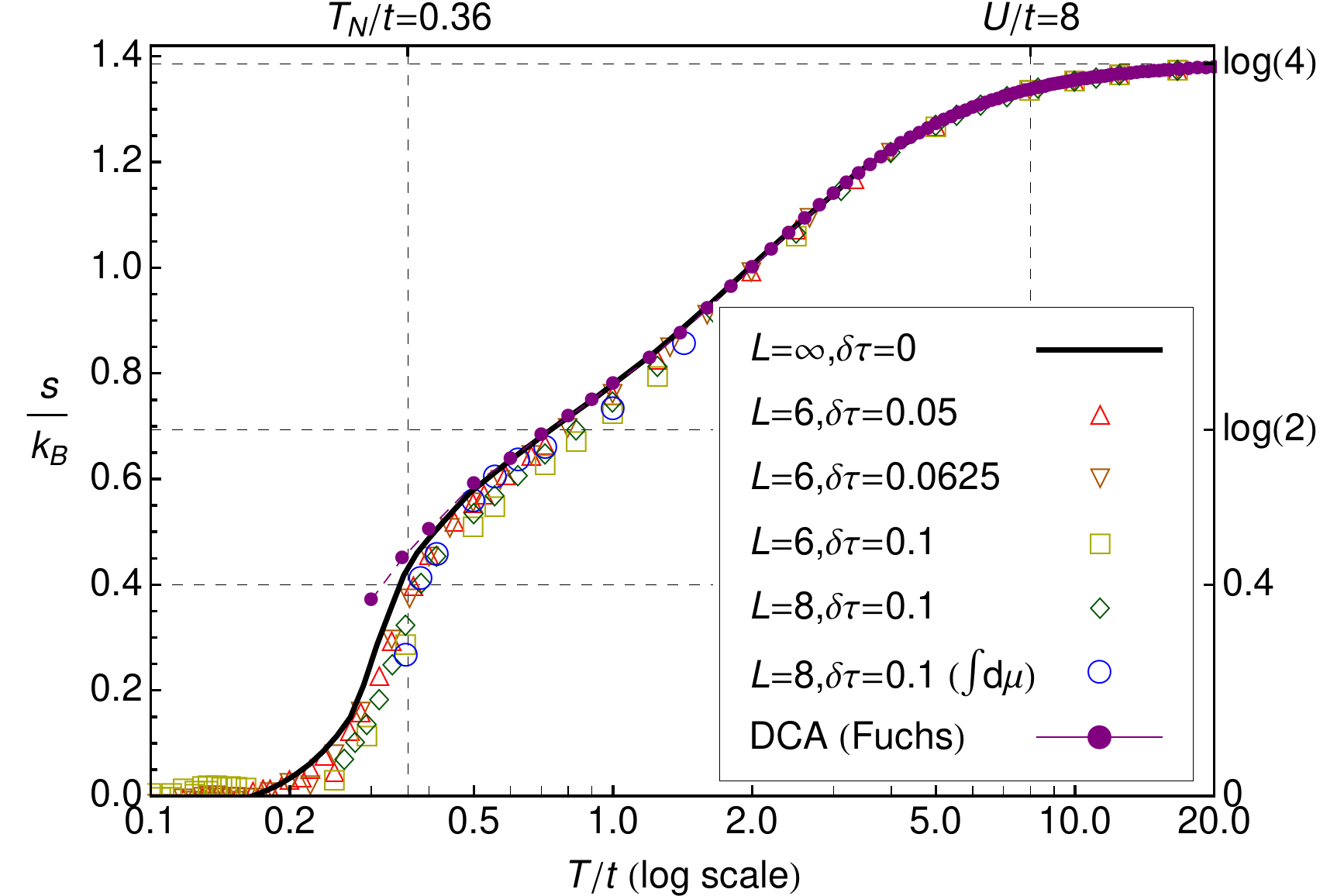}
		\label{sTFinal}
	}
	\vspace{5mm}
	\caption{
	\textbf{(a) Energy per site of homogeneous system}
	at half-filling and $U/t=8$,
	calculated using DQMC down to $T/t=0.1$.
	Statistical error bars are smaller than symbols.
	The solid curve is the entropy
		extrapolated to $L=\infty$ and $d\tau=0$
		(details in supplement).
	\textbf{(b) Entropy per site}
	obtained by integrating down from $T=\infty$,
	showing a shoulder at the Mott scale $T_\text{Mott} \simeq U$
	and a distinct feature at the N\'eel temperature $T_N \approx 0.36t$ due to critical fluctuations.
	Errors in $E/t$ and $s/k_B$ are both about $0.02$.
	DCA results \cite{fuchs2011} are shown for comparison.
	}
	\label{ETsTFinal}
	\end{figure}

In this Letter we present an adiabatic cooling protocol for trapped systems, 
which we expect to play an important role in the race for finding antiferromagnetism in the repulsive Hubbard model 
and for opening the door toward the search for the $d$-wave superfluid state.
We first calculate the thermodynamics of a homogeneous system using unbiased determinantal quantum Monte Carlo (DQMC)
as a function of filling and temperature, accessing both paramagnetic and AF phases.
At half-filling, this allows us to obtain the entropy down to $T=0.1t$ (see Fig.~\ref{sTFinal}),
well below the maximum N\'eel temperature $T_N \approx 0.36t$ \cite{staudt2000},
and also well below the temperatures accessed by recent cluster studies~\cite{fuchs2011}.

\begin{figure*}[!htb]
	\subfigure{
		\includegraphics[width=0.3\textwidth]{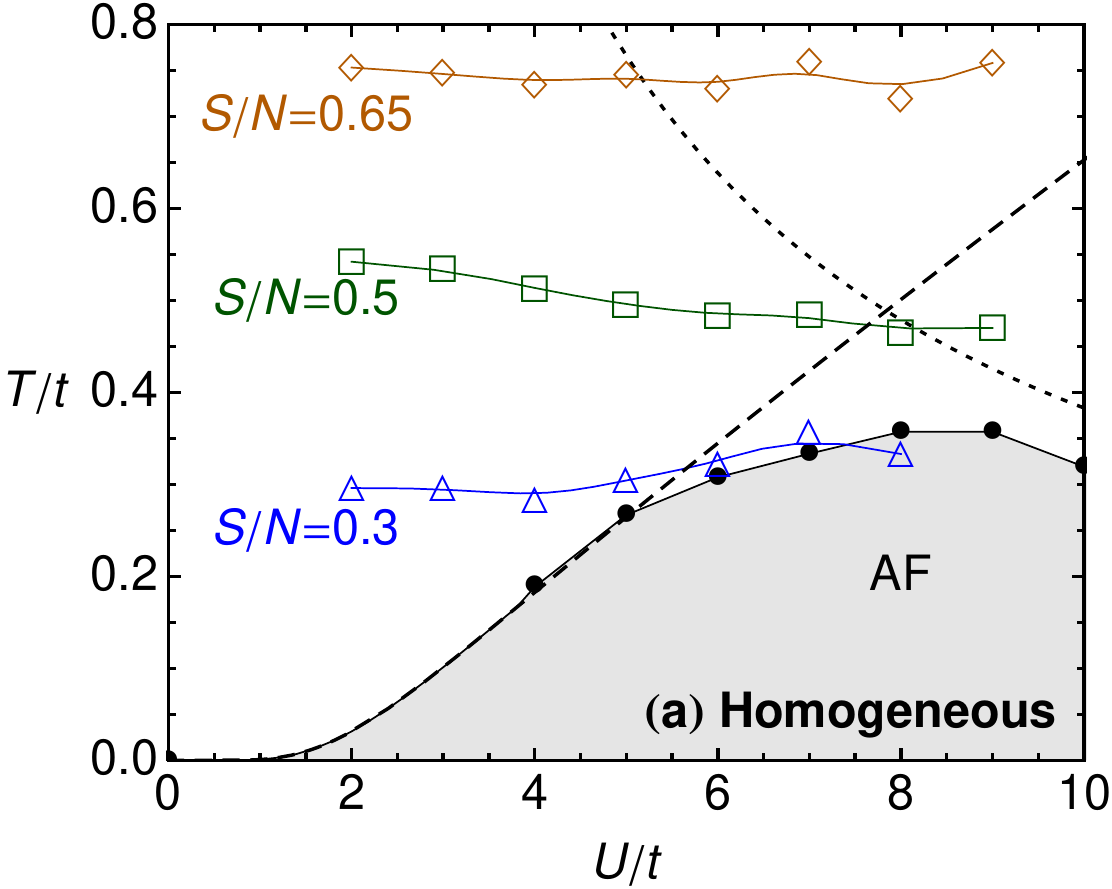}
		\label{HomogeneousIsentropes}
	}
	\subfigure{
		\includegraphics[width=0.3\textwidth]{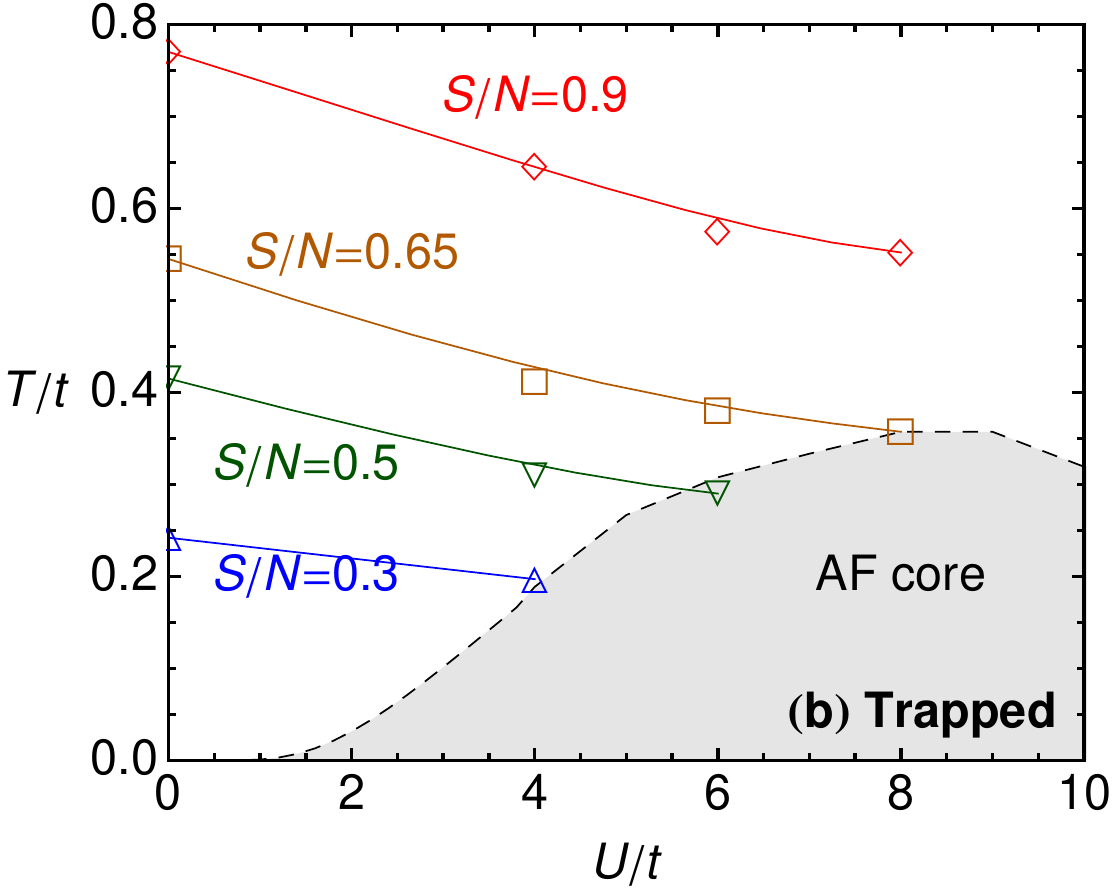}
		\label{TrappedIsentropes}
	}
	\subfigure{
		\includegraphics[width=0.3\textwidth]{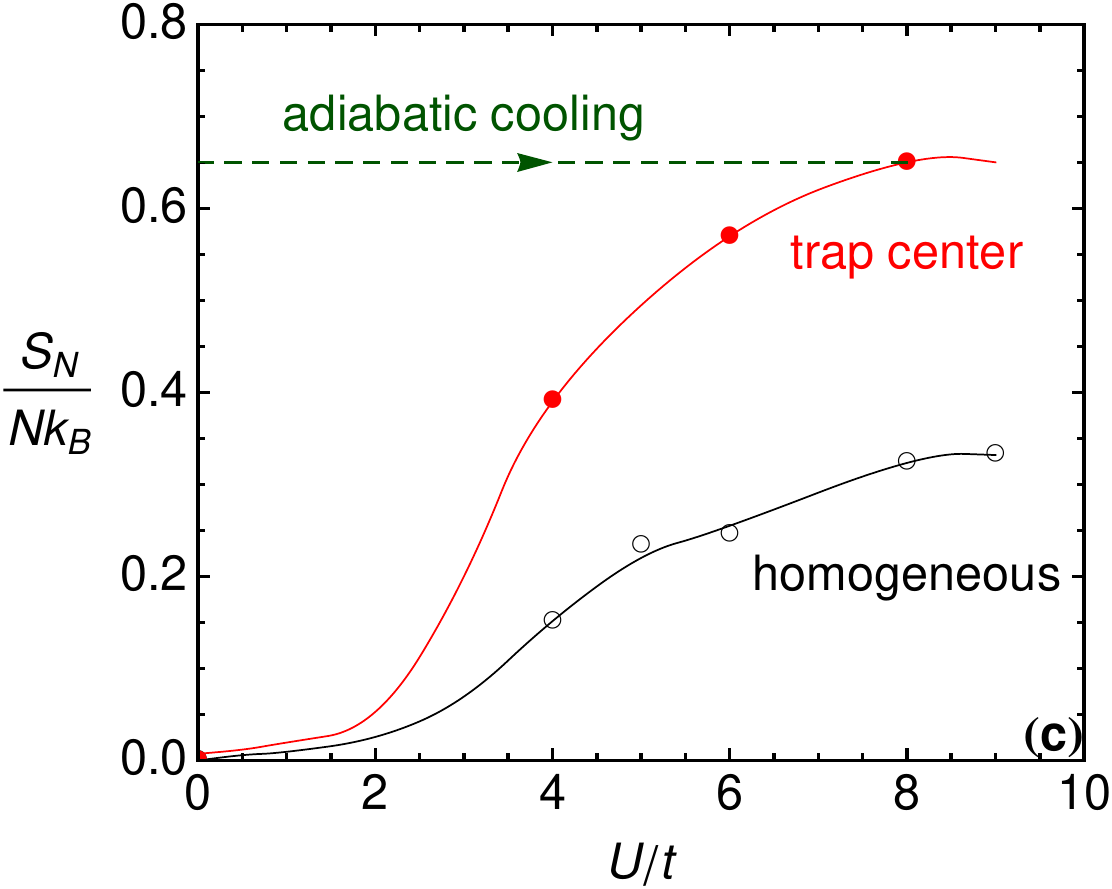}
		\label{sneel}
	}
	\caption{
	(a)
	Constant-entropy curves of a homogeneous system at half-filling~\cite{footnote1}.
	There is no clear evidence for ``Pomeranchuk'' cooling as $U$ is increased adiabatically,
		in marked contrast to (b).
	The filled symbols are QMC values for $T_N$ from Ref.~\onlinecite{staudt2000},
	and the dashed and dotted curves are weak- and strong-coupling asymptotic forms
	 (see text).
	(b)
	In a harmonic trap with $E_t=3.28t$,
		ramping up $U$ adiabatically
		produces significant cooling	due to entropy redistribution.
	An AF state can be produced in the trap center
		even for an overall entropy per particle $S/N\approx 0.65k_B$.
	(c)
	Average entropy per particle in a harmonic trap
		below which AF order exists at the center.
	This is significantly higher than the critical entropy of a homogeneous system.
	}
	\label{InteractionCoolingProtocol}
	\end{figure*}

\begin{figure*}[htb!]
	\includegraphics[width=0.32\textwidth,trim=25mm 5mm 30mm 20mm,clip]
		{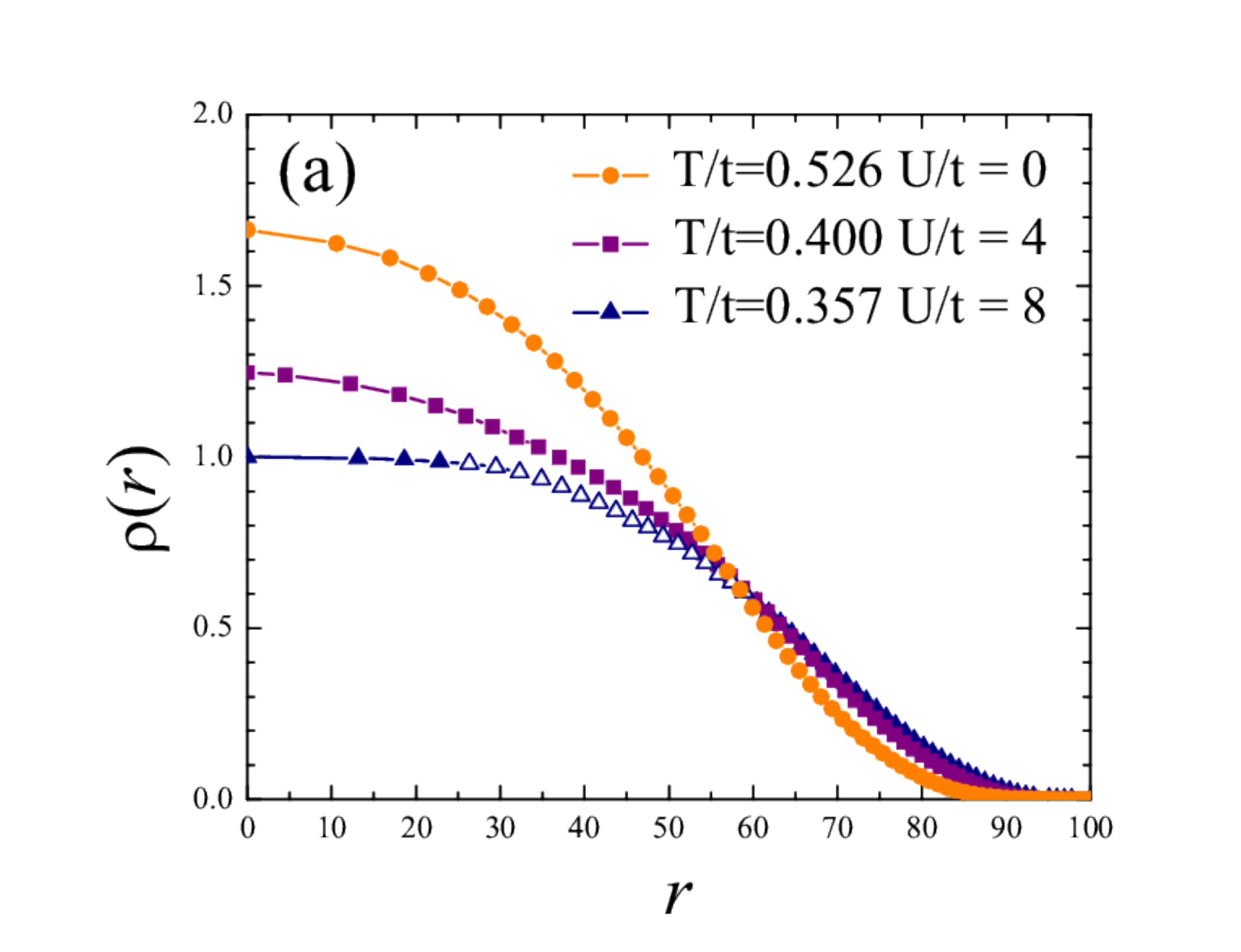}
	\includegraphics[	width=0.32\textwidth,trim=25mm 5mm 30mm 20mm,clip]
		{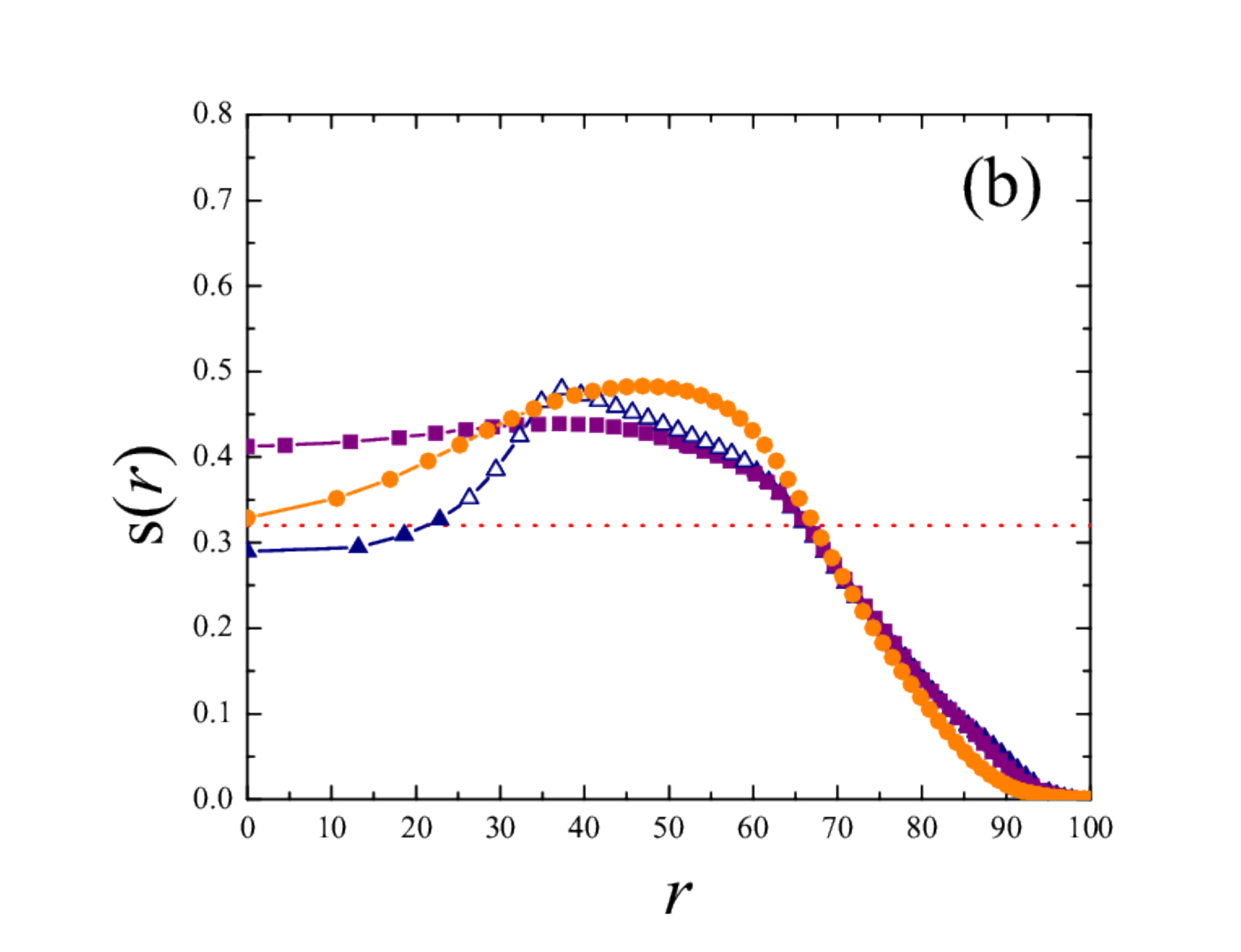}
	\includegraphics[	width=0.32\textwidth,trim=20mm 5mm 30mm 20mm,clip]
		{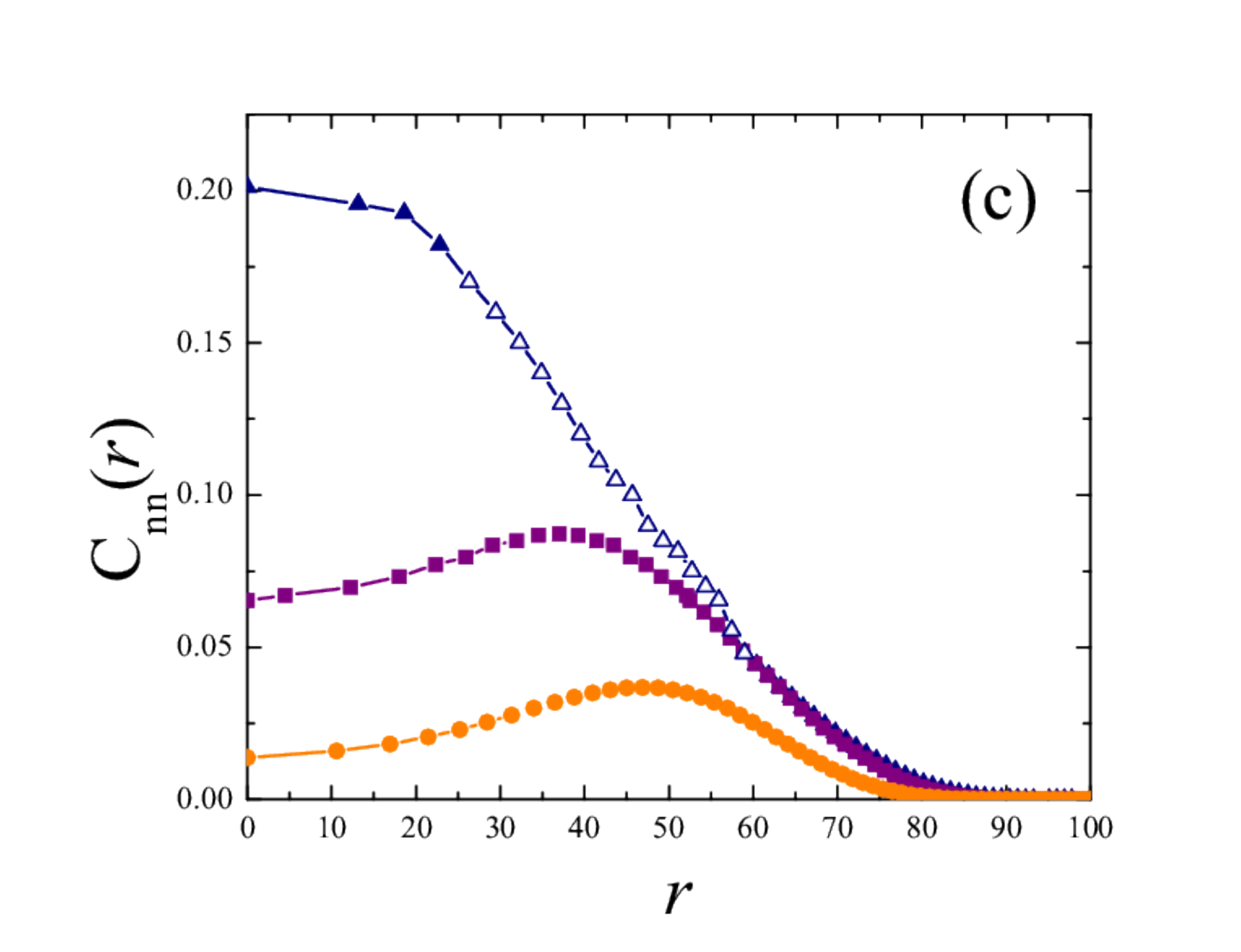}
	\vspace{-0.2cm}
	\caption{
	\textbf{Cooling by increasing interaction:}
	Adiabatic evolution of a cloud of $N=1.3\times 10^6$ particles
	with increasing interaction $U$
	for a fixed total entropy per particle $S/N=0.65k_B$
	and trap compression $E_t/t=3.28$.
	In going from $U/t=0$ to $8$, entropy is transferred from the core
	 to the wings (or outer shell) at $r>70$, 
	 	where $S = \int_0^\infty dr~ 4 \pi r^2 s(r) $ remains constant.
	\label{InteractionCoolingProfiles}
	}
	\end{figure*}

We next use the local density approximation to treat the effect of a harmonic trap.
We demonstrate that increasing the repulsion $U$ adiabatically leads to substantial cooling,
\emph{but only in the presence of the trap} (see Fig.~\ref{InteractionCoolingProtocol}).
During this process, the cloud expands and entropy gets redistributed from the center to the metallic wings. 
Even though the average entropy per particle $S/N\approx 0.65 k_B$ is higher than the critical entropy of the homogeneous system ($0.4k_B$ at $U/t=8$), it is nevertheless possible to generate an AF state at the center (see Fig.~\ref{InteractionCoolingProfiles}).

\myheading{Model and methods:}
We consider the Fermi-Hubbard Hamiltonian,
	\begin{eqnarray}
	\mathcal{H} = &-& t \sum_{\langle \rrr \rrr' \rangle \sigma}
	(\cdag_{\rrr\sigma} \cccc_{\rrr' \sigma} +
	\cdag_{\rrr\sigma} \cccc_{\rrr' \sigma} )
		+
	U\sum_\rrr n_{\rrr\uparrow} n_{\rrr\downarrow} 
		\nonumber \\
		&+& 
	\sum_\rrr (V_t\,\rrr^2 - \mu) (n_\rrr - 1) ,
	\label{hamiltonian}
	\end{eqnarray}
in which $\rrr$
labels a site (or well) of a 3D cubic optical lattice, $\sigma=\,\uparrow$ or $\downarrow$ corresponds to two
hyperfine states, $t$ is the nearest-neighbor hopping amplitude, $U$ is the on-site interaction energy,
$c_{\rrr \sigma}$ is the fermion destruction operator at site $\rrr$ with spin
$\sigma$, 
and
$n_{\rrr\sigma}=\cdag_{\rrr\sigma}c_{\rrr\sigma}$
 with $n_\rrr=\sum_\sigma n_{\rrr\sigma}$.  
The curvature $V_t=\frac{1}{2} m \omega_0^2 d^2$  describes harmonic confinement with trap frequency $\omega_0 / 2\pi$,
fermion mass $m$, and lattice spacing $d$.
The chemical potential $\mu$ controls the average density.
The parameters $t$ and $U$ can be directly related \cite{jaksch1998} to the lattice depth, set by the laser intensity, and to the interatomic interaction tuned by a Feshbach resonance. This Hamiltonian is valid in
the regime where only a single band is populated in the optical lattice. 
Following Ref.~\onlinecite{schneider2008} we define the characteristic trap energy $E_t = V_t (3N/8\pi)^{2/3}$.

We calculate the density $\rho$, energy density $E$, double occupancy 
$D=\langle n_{\rrr\uparrow} n_{\rrr\downarrow} \rangle$, and spin correlations for a homogeneous system ($V_t=0$) 
as a function of $\mu$, $T$, and $U/t$ 
using DQMC simulations~\cite{blankenbecler1981,white1989}.

\myheading{Half-filling:}
We first focus on the homogeneous case at half-filling ($\mu=0$) and $U/t=8$,
where the N\'eel temperature $T_N/t = 0.36$ is highest~\cite{staudt2000}. 
At $\mu=0$ DQMC is free of the fermion sign problem and we can access low temperatures down to $T=0.1t$, well into the AF phase.
We perform extrapolation on $E(T)$ to the limit of zero imaginary-time discretization ($\delta\tau=0$) and infinite system size ($L^3=\infty$), as described in detail in the supplement.  
The high statistical accuracy of the DQMC data even reveals critical fluctuations near $T_N$.

We obtain the ground state energy $E_0/t = -0.74(2)$ and the correct low-temperature behavior ($E \sim T^4$) expected for an antiferromagnet with linearly dispersing spin waves.  
The results are shown in Fig.~\ref{ETFinal}.
Integrating $E(T)$ down from infinite temperature, we determine the entropy per site using $s(T) = \ln 4 + E/T - \int_T^\infty dT~ E/T^2$.
Our results agree with extrapolated results from the dynamical cluster approximation (DCA)~\cite{fuchs2011}, available only in the paramagnetic phase.

We see from Fig.~\ref{sTFinal} that as the temperature is reduced below $U=8t$, 
the entropy per site $s/k_B$ decreases from $\ln(4)$ to $\ln(2)$,
due to suppression of double occupancy below the Mott scale for charge fluctuations.
At $T_N$ the critical entropy is $s_N/k_B \approx 0.4 k_B$.
Our DQMC results show a steep drop in entropy below $T_N$ resulting from spin ordering.

In Fig.~\ref{HomogeneousIsentropes}, we show constant-entropy curves in the $(T,U)$ plane at half-filling.  We also plot the N\'eel temperature as a function of $U$ obtained from previous QMC simulations~\cite{staudt2000} together with its asymptotic forms at weak and strong coupling.
The dashed curve is $0.282 T_\text{MF} (U/t)$ where the mean-field result is given by
$
	2/U
	= \sum_k \tanh (2\epsilon_k / T_\text{MF}) / \epsilon_k
$
and the suppression factor 0.282 arises from $O((U/t)^2)$ vertex corrections~\cite{gmb1961,vandongen1994}.
The dotted curve shows the strong-coupling Heisenberg limit result $3.78t^2/U$~\cite{sandvik1998}.

\myheading{Away from half-filling:}
We next compute the equation of state $\rho(\mu)$ of the homogeneous system away from half-filling, as this will be needed to study the effect of a trap.
We now obtain the entropy by integrating along an isotherm from the empty lattice,
$	s(\mu)
	= \int_{-\infty}^{\mu} d\mu~
	(\partial s/\partial \mu) _T
$,
making use of the Maxwell relation
$(\partial s/\partial \mu)  _T
	=
	(\partial \rho/\partial T)  _\mu
$,
where $(\partial \rho/\partial T)  _\mu$ 
is evaluated using a finite difference scheme.
This gives results (indicated by symbols labelled ``$\int d\mu$'' in Fig.~\ref{sTFinal}) consistent with integration of $E(T)$ as described above.

\begin{figure*}[htb!]
	\includegraphics[width=0.3\textwidth,trim=20mm 5mm 30mm 20mm,clip]{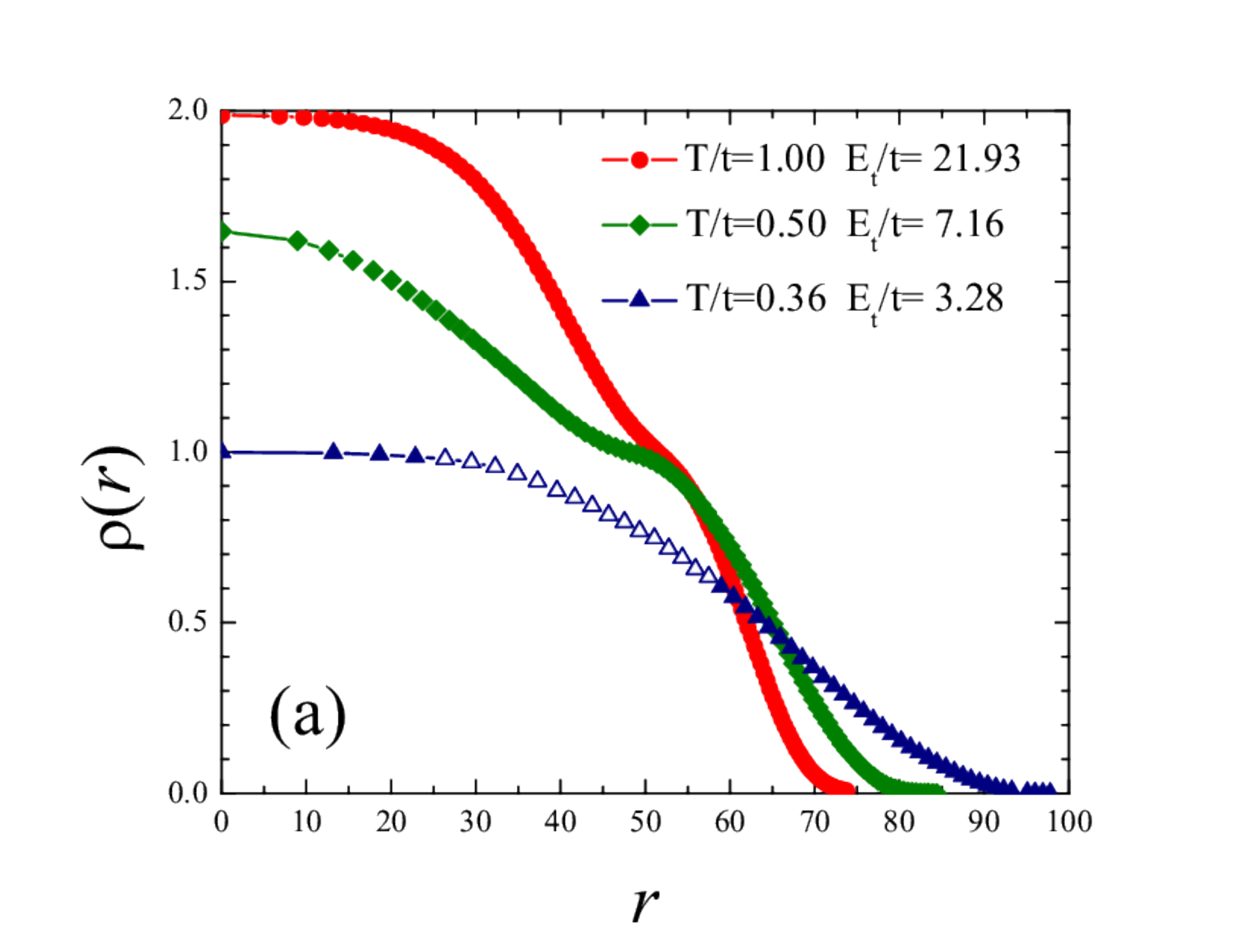}
	\includegraphics[width=0.3\textwidth,trim=20mm 5mm 30mm 20mm,clip]{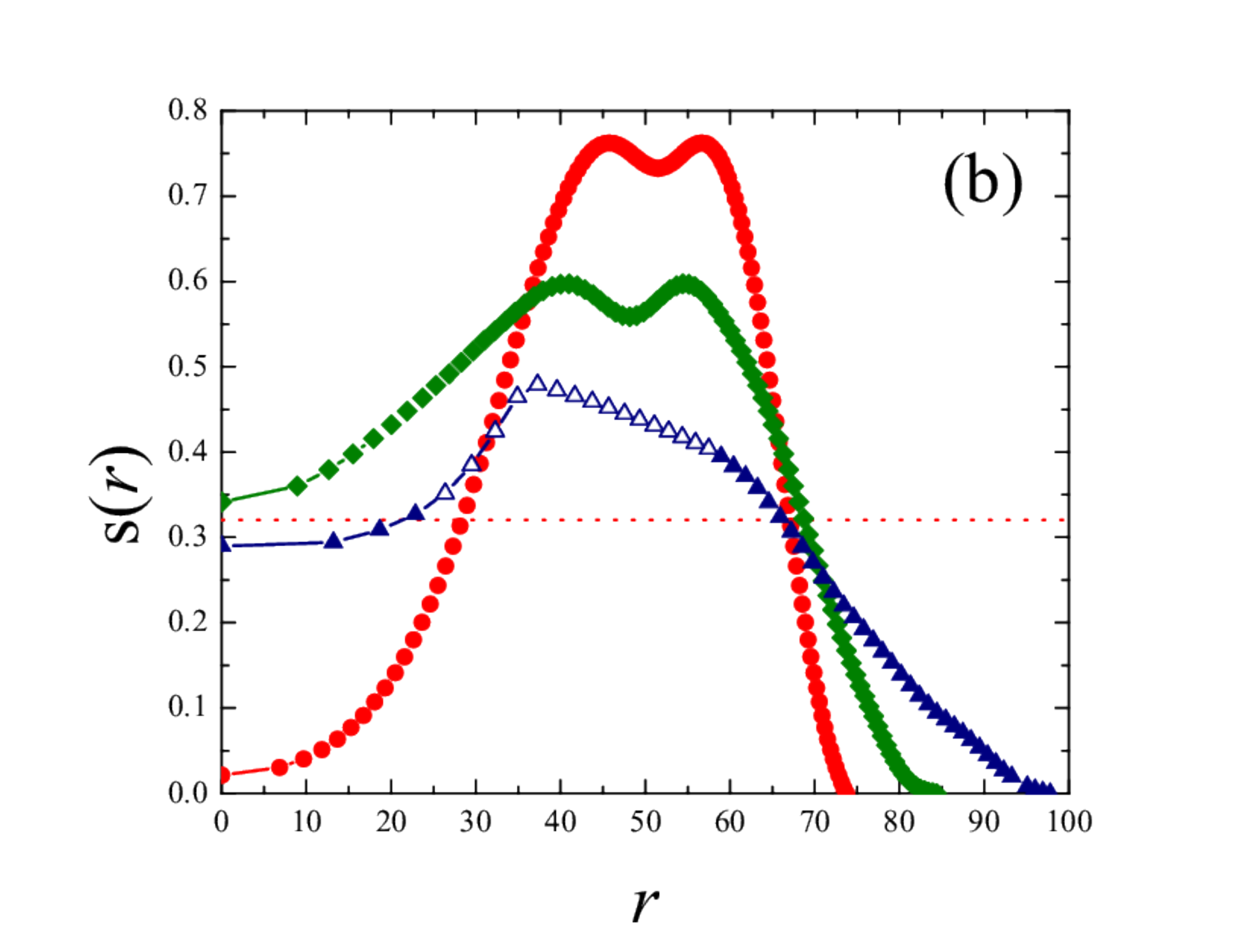}
	\includegraphics[width=0.3\textwidth,trim=5mm 5mm 30mm 20mm,clip]{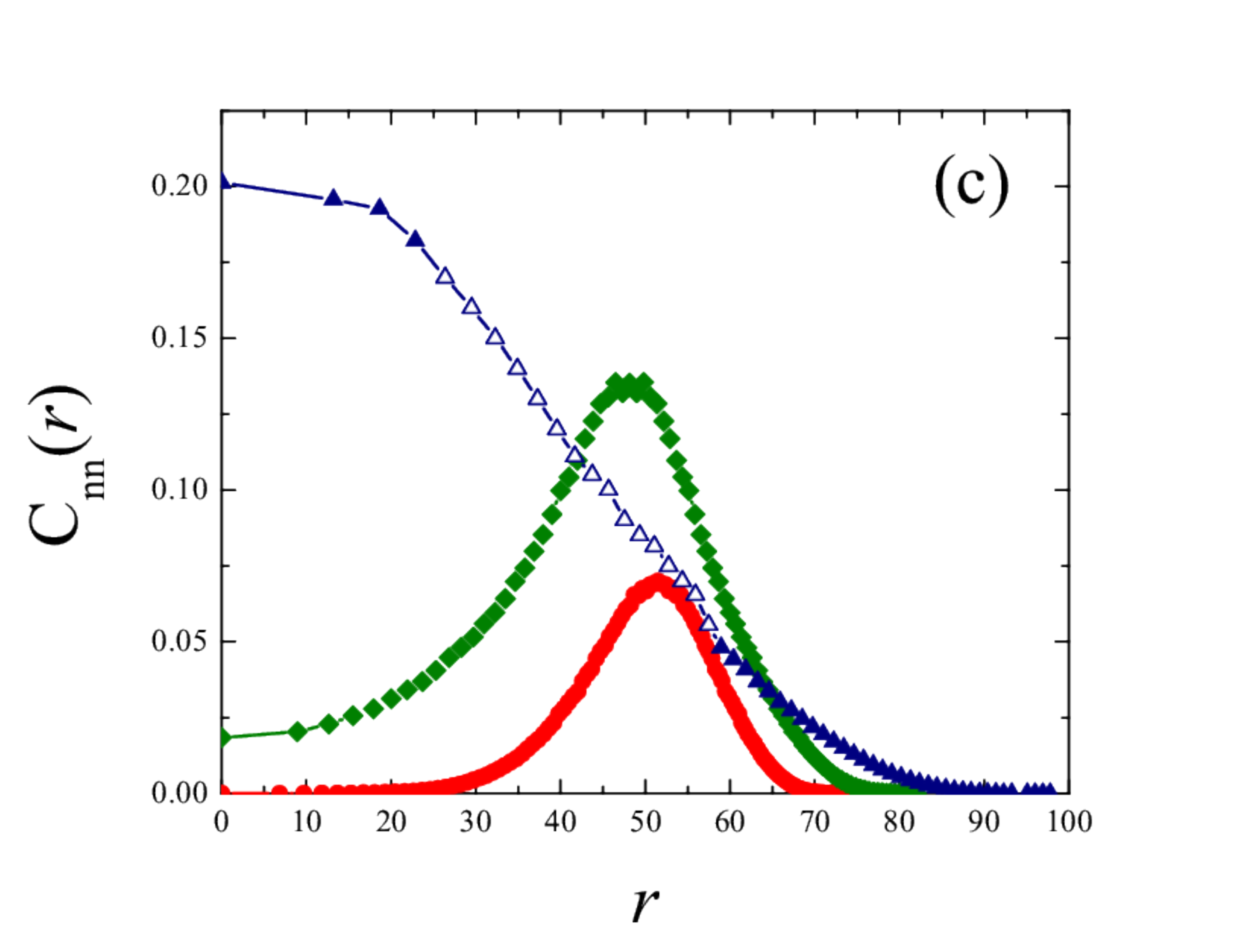}
	\vspace{-0.2cm}
	\caption{
	\textbf{Cooling by expansion:}
	Adiabatic evolution of a cloud of $N=1.3\times 10^6$ particles
		with decreasing trap energy $E_t$
		for a fixed total entropy per particle $S/N=0.65k_B$ 
		at interaction $U/t=8$.
	}
	\label{ExpansionCoolingProfiles}
	\end{figure*}

We model the trap using the local density approximation (LDA),
in which local observables are given by their homogeneous values evaluated at a chemical potential
$\mu(r) = \mu_0 - V_t \rrr^2$.
The chemical potential at the trap center $\mu_0$ is determined from the total fermion number $N = \int_0^\infty dr~  4 \pi r^2 \rho(\mu(r))$.
We obtain density, entropy, and local spin correlation profiles such as those in Figs.~\ref{InteractionCoolingProfiles} and \ref{ExpansionCoolingProfiles}, from which we can deduce a route to achieving cooling in optical lattices.

\myheading{Cooling:}
Note the contrast between the constant-entropy curves in the homogeneous system at half-filling
(Fig.~\ref{HomogeneousIsentropes})
and in a harmonic trap with $E_t=3.28t$
(Fig.~\ref{TrappedIsentropes}).
For a given entropy per particle $S/N$ the temperature of the trapped system is already lower than that of the homogeneous system at $U=0$.  Furthermore, as $U$ is ramped up, the trapped system exhibits significant cooling compared to the homogeneous system.
Thus we see that for $E_t=3.28t$ and any starting entropy less than $0.65k_B$, 
one can obtain an AF core by adiabatic cooling (see Fig.~\ref{sneel}).

We gain further insight from the profiles shown in Fig.~\ref{InteractionCoolingProfiles}(a,b,c).
As the interaction is ramped up from $U/t=0$ to $8$,
the cloud expands and the density at the center decreases towards 1, characteristic of a Mott insulator (MI).  This MI has a gap to charge excitations and thus a low entropy.
On the other hand, the metallic state in the wings, with its low-energy spin and charge excitations,
can act as an entropy sink.
Thus, 
entropy is transferred from the Mott core to the metallic wings.

During this process the temperature falls from $T/t=0.53$ to $0.36 \approx T_N$.
In the final state, the entropy density $s(r)$ at the center is near the critical value for AF ordering indicated by the dashed line~\cite{footnote1}.  We see the growth of local antiferromagnetic correlations from the nearest-neighbor spin-spin correlation
$C_{nn}(r) = - \langle \SSS_\rrr \cdot \SSS_{\rrr+\hat{\mathbf{x}}}\rangle$,
where
	$\SSS_\rrr = \frac{1}{2} 
	\sum_{\alpha\beta} \pmb{\sigma}_{\alpha\beta} \cdag_{\rrr\alpha} \cccc_{\rrr\beta}
	$ 
is the spin at site $\rrr$.

Our analysis shows that the adiabatic cooling in a trap results from entropy redistribution, and not from a Pomeranchuk effect in the homogeneous equation of state~\cite{werner2005,deleo2011} as discussed below.  In any case, we do not find a significant Pomeranchuk effect $(\partial T/\partial U)_S < 0$  in DQMC, either in 3D (see Fig.~\ref{HomogeneousIsentropes}) or in 2D~\cite{paiva2010,dare2007}.

Another way to cool in a trap is to use adiabatic expansion, a standard cryogenic technique, the results for which are shown in Fig.~\ref{ExpansionCoolingProfiles}.
We see that as $E_t/t$ decreases from 21.93 to 3.28, the core goes from a band insulator to an antiferromagnetic MI. 

In Figs.~\ref{InteractionCoolingProfiles} and \ref{ExpansionCoolingProfiles}
the open symbols 
used only at the lowest temperature ($T/t=0.36t$)
denote regions of the trap away from half-filling 
where the DQMC sign problem is significant.
In this range we have used a combination of interpolation and results from smaller systems (for which the sign problem is less severe).

	\begin{figure}[htb!]
	\subfigure[]{
		\includegraphics[width=0.99\columnwidth,trim=10mm 5mm 30mm 20mm,clip]
			{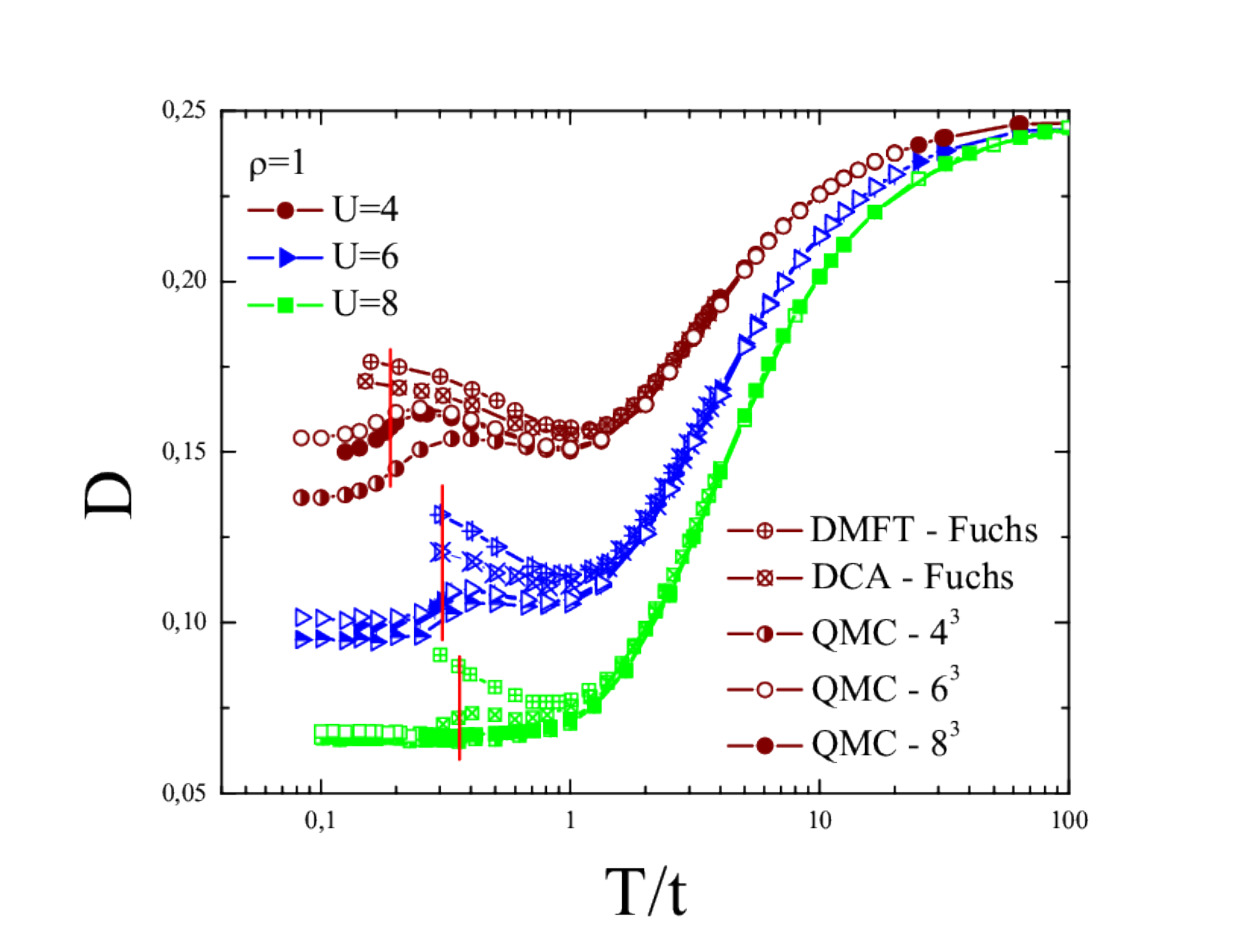}
	}
	\vspace{-0.2cm}
	\caption{
		\textbf{Double occupancy $D(T)$}
		for the homogeneous system at half-filling	for
			$U/t=4, \delta\tau=0.125$;
			$U/t=6, \delta\tau=0.125$; and
			$U/t=8, \delta\tau=0.1$.
		The vertical red lines correspond to $T_N(U)$.
		\label{DFuchs}
	}
	\end{figure}
We now remark on the temperature dependence of the double occupancy $D$ of the homogeneous system at half-filling, shown in Fig.~\ref{DFuchs}.  
As $T$ decreases below the $U$, $D$ is generally suppressed due to Mott physics, so that $(\partial D/\partial T)_U > 0$.
 At low temperature for intermediate $U/t=4$ to $6$, $D$ shows anomalous behavior in that $(\partial D/\partial T)_U < 0$ over a range of $T$ close to $T_N$.
Using a Maxwell relation,
$(\partial D/\partial T)_U 
= (\partial D/\partial S)_U (\partial S/\partial T)_U
= (\partial T/\partial U)_S \times C / T
$,
so that
$(\partial T/\partial U)_S < 0$,
suggesting the possibility of ``Pomeranchuk cooling''~\cite{werner2005} by adiabatically increasing the interaction.
However, the effect is smaller than predicted by DMFT and DCA~\cite{fuchs2011}.
This supports our conclusion that the ``Pomeranchuk effect'' in a homogeneous system is insignificant, as already shown in Fig.~\ref{HomogeneousIsentropes}.

\myheading{Discussion and conclusion:}

To conclude, our most significant observation is that it is possible to lower the temperature of the 
{\em trapped} system by suitable adiabatic processes.
Cooling results from entropy redistribution in a trap with the metallic wings acting as entropy sinks.
We find that 
an average entropy per particle in the trap $S/N=0.65 k_B$
is sufficiently low to produce an AF state at the center
using our adiabatic cooling protocol.
In order to go well below $T_N$ a correspondingly lower entropy is required.

The results for the trapped system are markedly different from those for the homogeneous system.
First, the maximum critical entropy of a homogeneous AF state occurring at $U=8t$ 
is $0.4 k_B$, considerably lower than the average value required in a trap.
Second, adiabatically increasing $U$ in the homogeneous case does not lead to significant cooling.

We finally discuss the implications for optical lattice experiments~\cite{jordens2008,schneider2008}.
Before the lattice is turned on, the initial temperature of a trapped gas is typically 
$T_i \sim 0.1 T_F$
where the Fermi temperature $k_B T_F=\hbar\omega_0 (3N)^{1/3}$.
For non-interacting fermions, an initial temperature $T_i/T_F \approx 0.06$,
within the reach of current experiments,
corresponds to an average entropy per particle $S/N=0.65 k_B$ in the trap.
As noted above, this leads to an AF state at the center,
which can be probed by the growth of nearest-neighbor spin-spin correlations.
Thus, the results presented here imply that antiferromagnetism is achievable in optical lattices, provided that adiabaticity can be maintained during our cooling protocol.

We gratefully acknowledge support from 
FAPERJ, CNPq, and INCT on Quantum Information (TCLP),
DARPA grant no. W911NF-08-1-0338 (YLL),
ARO W911NF-08-1-0338 (MR,NT),
ARO W911NF-07-1-0576 (RTS), 
and the DARPA OLE Program (RTS,NT).       
We acknowledge fruitful discussions with U. Schneider. 
RTS is grateful to Thomas Maier and Mark Jarrell for discussions regarding the comparison of DQMC with DCA.


\begin{thebibliography}{10}
\bibitem{footnote1}
For the results in a trap we used a DQMC equation of state obtained from
  $L^3=8^3$ and $\delta\tau=0.1$, with a critical entropy $s_N \approx 0.32
  k_B$ at half-filling. While this slightly underestimates the entropy, the
  qualitative conclusions regarding entropy redistribution and cooling in a
  trap are not affected.
\end{thebibliography}

\end{document}


\title{Supplemental Information for ``Fermions in 3D Optical Lattices: Cooling Protocol to Obtain Antiferromagnetism''}
\author{
Thereza Paiva$^{1}$, 
Yen Lee Loh$^2$, 
Mohit Randeria$^{2}$,
Richard T. Scalettar$^3$,
and
Nandini Trivedi$^2$
}
\affiliation{
Instituto de Fisica, Universidade Federal do Rio de
Janeiro Cx.P. 68.528, 21941-972 Rio de Janeiro RJ, Brazil\\
$^{2}$Department of Physics, The Ohio State University, Columbus, OH
43210, USA \\
$^{3}$Department of Physics, University of California, Davis, CA 95616, USA
}
\maketitle

\myheading{Determinantal quantum Monte Carlo:}
We calculate $\rho(\mu,T,U/t)$ in the homogeneous case (no trap) using
determinantal quantum Monte Carlo(DQMC) \cite{blankenbecler1981,white1989}. 
The DQMC approach provides an exact solution on finite spatial lattices
by retaining the full space and imaginary-time dependence of the
fermion Green function.  This is accomplished through
the introduction of an auxiliary field which allows for the
trace over the fermionic degrees of freedom to be done analytically,
the sum over the auxiliary field then being performed stochastically.
At half-filling $\mu=0$ there is no sign problem and the spin correlations can be calculated down to low temperatures.
However, in a trap the density becomes inhomogeneous and in order to capture that behavior we need the density
away from half-filling. These calculations are difficult since the probability weights being used to generate the configurations in the QMC algorithm can become negative.

%
%
%

	\begin{figure*}[htb!]
	\subfigure[]{
		\includegraphics[width=0.99\columnwidth]{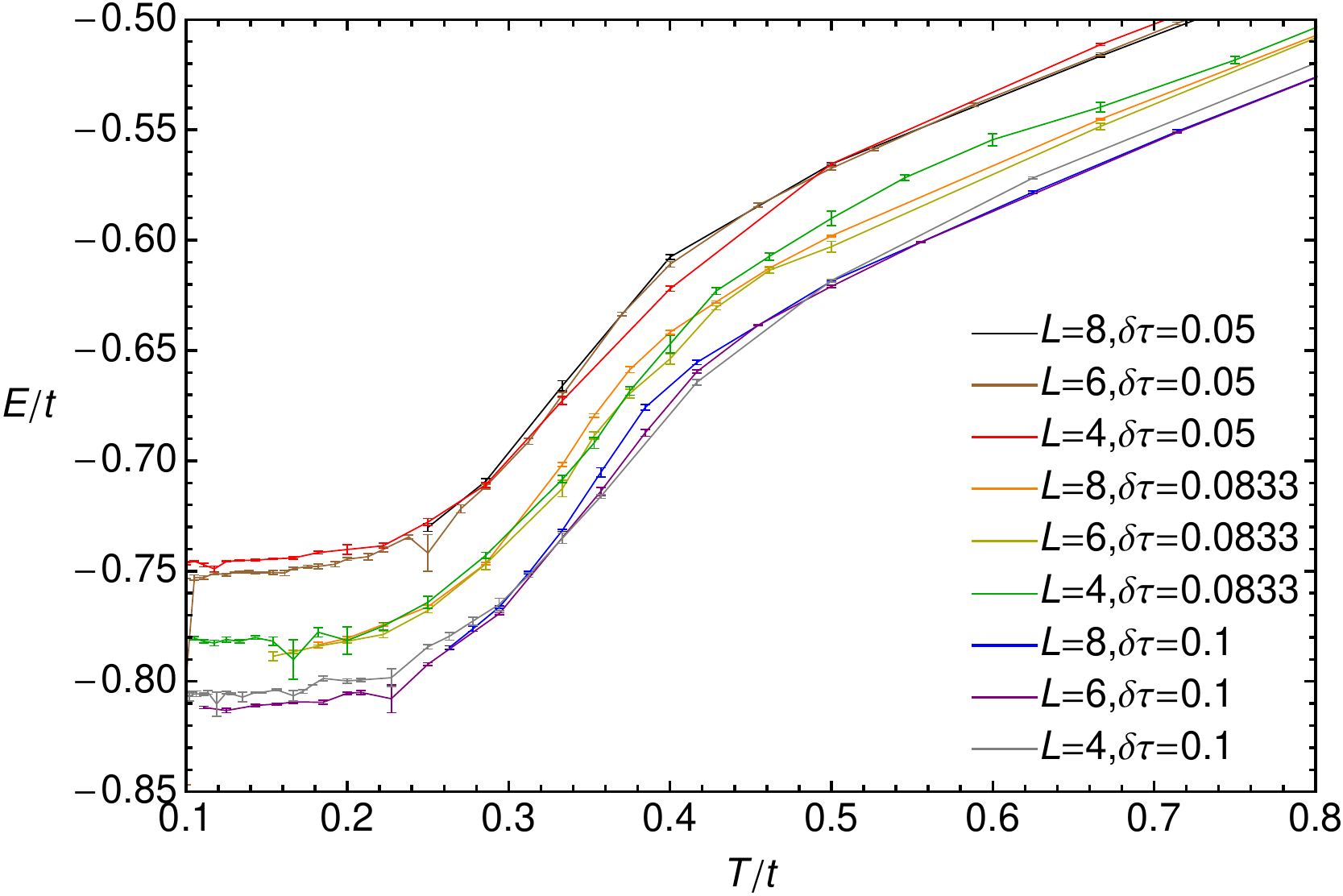}
		\label{scaling1}
	}
	\subfigure[]{
		\includegraphics[width=0.99\columnwidth]{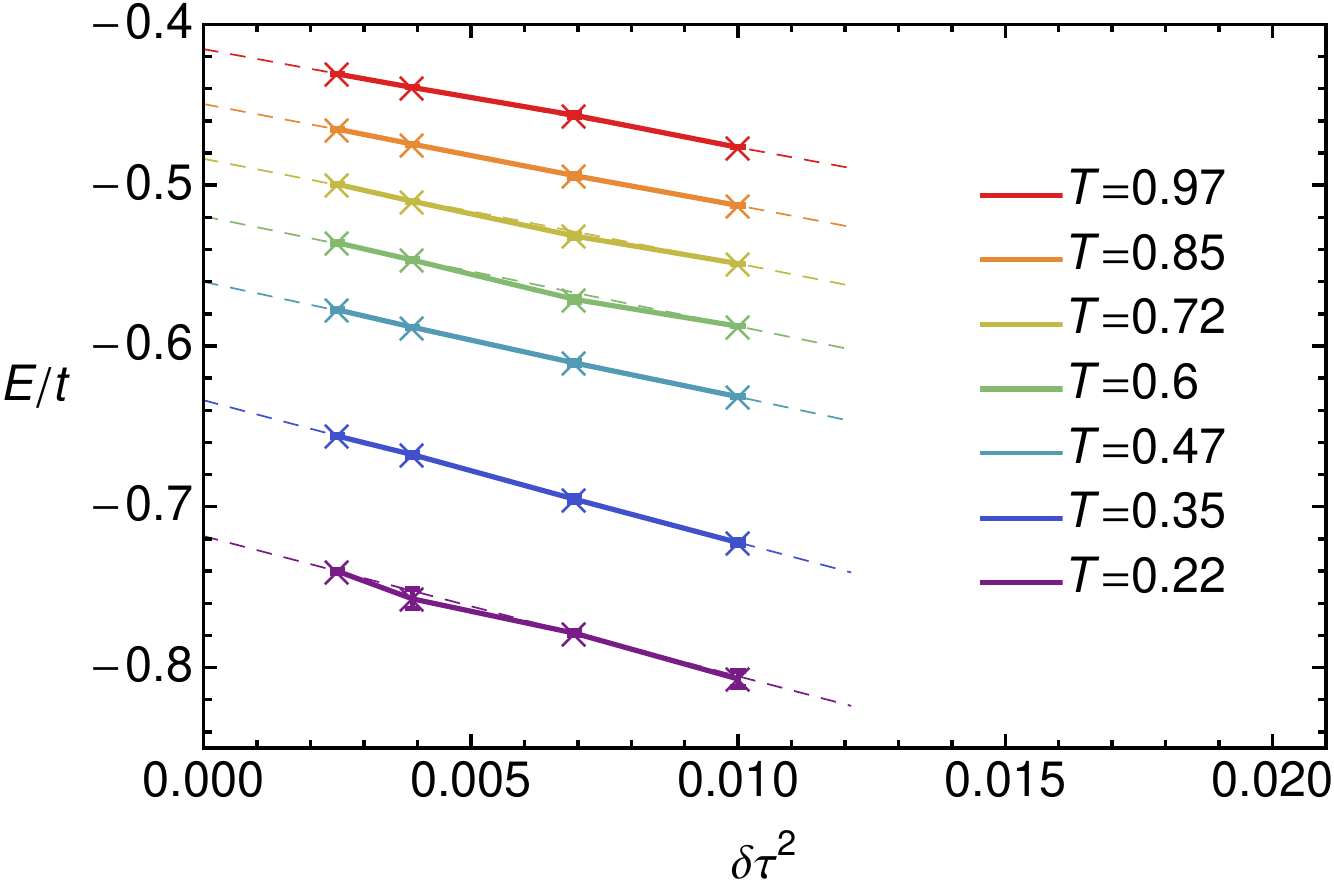}
		\label{scaling2}
	}
	\subfigure[]{
		\includegraphics[width=0.99\columnwidth]{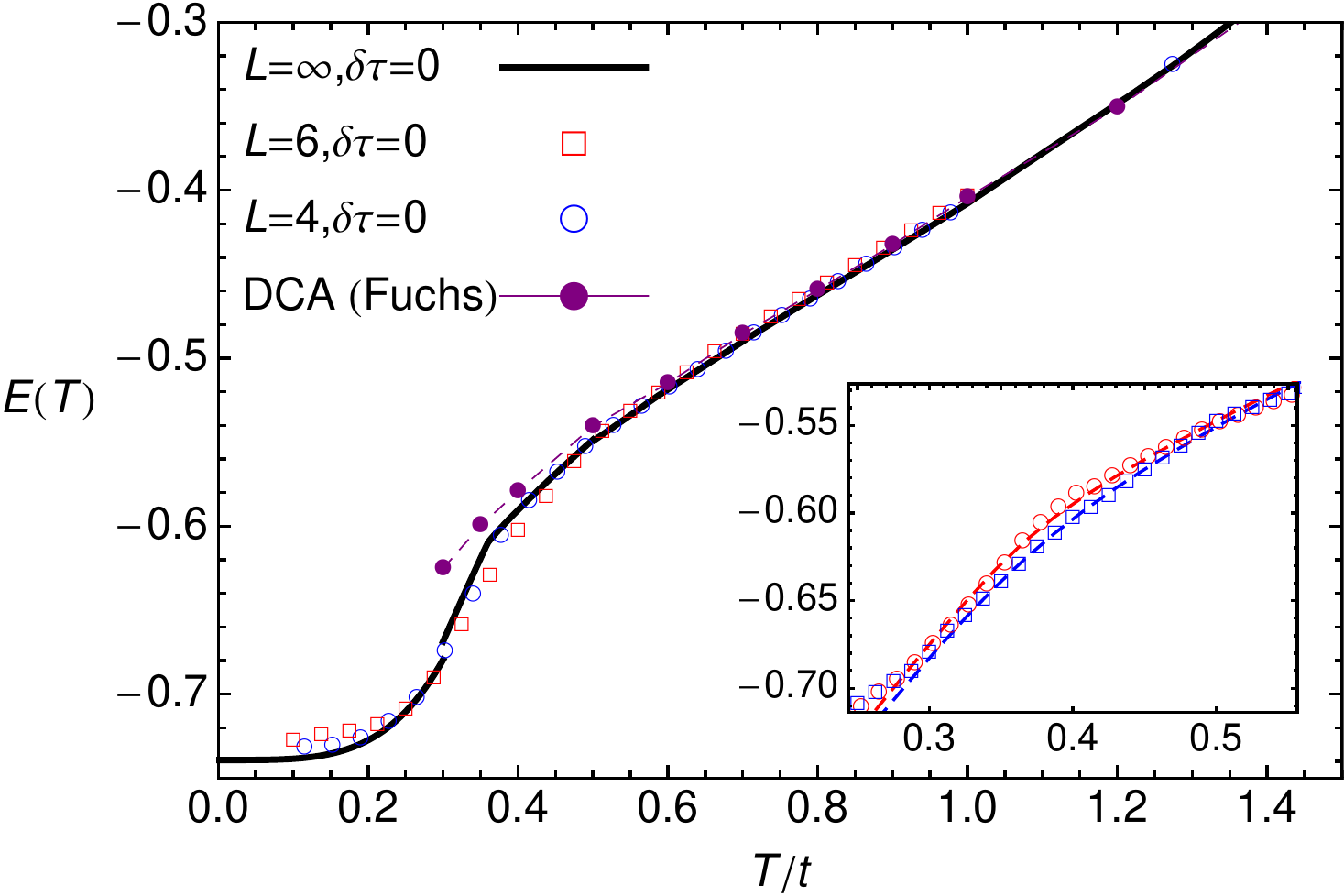}
		\label{scaling4}
	}
	\vspace{-0.2cm}
	\caption{
	\textbf{Elimination of $\delta\tau$- and finite-size errors:}	
	(a) Energy per site $E(T)$ for various system sizes $L^3$ 
		and imaginary time discretization steps $\delta\tau$.
	(b) $E(\delta\tau)$ for $L=6$ at various $T$, showing $\delta\tau^2$ scaling.
		Most error bars are smaller than the linewidths.
	(c) $E(T)$ extrapolated to $\delta\tau=0$ and $L=\infty$
		using suitable fitting forms at low, medium, and high temperatures.
		The inset shows a close-up;
			dashed red and blue curves are fits with scaling forms
			near criticality (see text).			
		The accuracy of the extrapolated energy is about $0.02t$.
		Extrapolated DCA results from Ref.~\onlinecite{fuchs2011} are shown for comparison.			The extrapolated QMC curve (black) captures the critical fluctuations near the 
		N\'eel temperature and falls below the DCA results.
	}
	\label{TrotterScaling}
	\end{figure*}

The QMC data is of sufficient quality that we can identify finite-size and finite-$\delta\tau$ effects and perform an extrapolation to $L^d = \infty$ and $\delta\tau = 0$.
We have carried out this procedure for $U/t=8$ and $\mu=0$.

The finite-$\delta\tau$ errors in $E(T)$ are about $0.05t$ (see Fig.~\ref{scaling1}), but because they involve non-critical high-energy degrees of freedom, they simply scale as $\delta\tau^2$ (as expected from Ref.~\onlinecite{trotterscaling}) and can be eliminated by extrapolation (see Fig.~\ref{scaling2}).
The finite-size errors are smaller, of the order of $0.02t$ (see inset of Fig.~\ref{scaling4}), 
but the discrepancy between $L=4$ and $L=6$ curves for $0.3 < T/t < 0.5$ is a clear signature of critical behavior, which demands a more sophisticated treatment.
Thus, in this temperature range, we fit multiple curves $E(T,L)$ simultaneously using an appropriate scaling ansatz involving the known critical exponents and amplitude ratios ($\nu \approx 0.70$, $\alpha = 2-d\nu = -0.11$, $A \equiv A_+/A_- \approx 1.52$) of the 3D Heisenberg universality class~\cite{chaikin}, using the critical temperature from the literature \cite{staudt2000}, $T_N/t \approx 0.36$, and including a non-singular background.
This allows extrapolation to the infinite-size limit, in which $E(T)$ has a kink at $T_N$ (see Fig.~\ref{scaling4}).
At low temperature we extrapolate $E(T,L)$ to $L=\infty$ assuming $1/L^2$ scaling, and the result is well fit by a power law due to linearly dispersing antiferromagnons, $E(T) = E_0 + c_1 T^4$.
At high temperature finite-size errors are negligible and $E(T)$ matches the high-temperature series expansion from Ref.~\onlinecite{fuchs2011}.

DQMC is a complementary approach to the dynamical cluster approximation (DCA).
DQMC gives ``exact'' results for an isolated cluster (after Trotter extrapolation),
whereas DCA simulates a cluster in a self-consistent bath~\cite{fuchs2011}.  
Both require finite-size scaling.
The present DQMC calculations have succeeded in going to low temperatures even below $T_N$ and densities away from half-filling.
In contrast, the present implementation of DCA does not include an order parameter, which precludes its use below $T_N$.
Our scaling analysis suggests a critical entropy $s_N \approx 0.4$, which is consistent with Ref.~\onlinecite{fuchs2011} to within uncertainties.

\myheading{Thermodynamic relations:}
From DQMC, one can calculate the density with respect to half-filling $\trho = \rho - 1$, energy per site $E$, and average double occupancy $D$ as functions of $\mu$, $T$, and $U$.  From these quantities it is possible to construct the grand potential per site $\Omega(\mu,T,U)$ in multiple ways, for example:
	\begin{align}
	\Omega(\mu,T) &= \frac{U}{4} - \mu + \int_\mu^\infty d\mu~ \trho
	;
	\\
	\Omega(\mu,T) &= -T \ln 4 + T \int_0^\beta d\beta~ (E - \mu \trho)
	.
	\end{align}
One can thence derive formulas for all other thermodynamic functions, such as the entropy per site:
	\begin{align}
	s(\mu,T) &= 0 
		+ \int_{-\infty}^{\mu} d\mu~	\frac{\partial \rho}{\partial T}  \bigg|_\mu
	;
	\\
	s(\mu,T) &=
		\ln 4 + \beta (E - \mu \trho) - \int_0^\beta d\beta~ (E - \mu \trho) 
	.
	\end{align}

\myheading{Justification for LDA:}
The data presented here are obtained by simulating
homogeneous ($V_t=0$) lattices, 
and then obtaining results for the confined system using the local
density approximation (LDA).  
Past comparisons with DQMC 
simulations with $V_t \neq 0$ suggest
that the LDA provides high accuracy
for the local profiles of the filling, spin, compressibility, and
entropy even in 2D \cite{chiesa2011}, and hence should
be even more accurate here in 3D.
